\documentclass[12pt, letterpaper]{article}

\usepackage[utf8]{inputenc}
\usepackage[superscript]{cite}
\usepackage[margin=1in, top=0.5in]{geometry}
\usepackage{setspace}
\usepackage{cite}
\usepackage{graphicx}
\usepackage{authblk} % Standard for handling multiple authors/affiliations
\usepackage{abstract}
\usepackage{authblk}
\usepackage{xspace}

\newcommand{\VBm}{\ensuremath{\mathrm{V}_{\mathrm{B}}^{-}}\xspace}
\title{\textbf{Symmetry Engineering Enables Deterministic Ensemble Polarization Control in Hexagonal Boron Nitride}}
\usepackage{authblk}

\author[1]{Aqiq Ishraq}
\author[1]{Eric Herrmann}
\author[1]{Pragya Agnihotri}
\author[1,2]{Alexander Hutchinson}
\author[1]{Ramiro M. dos Santos}
\author[1]{Shahidul Asif}
\author[1]{Lottie Murray}
\author[1]{Abhijith Puthiya Veettil}
\author[1]{Luke Stockl}
\author[2]{Muhammad Hassan Shaikh}
\author[1]{Collin Maurtua}
\author[4]{Kenji Watanabe}
\author[4]{Takashi Taniguchi}
\author[1,3]{Matthew Doty}
\author[5,6]{Cyrus E. Dreyer}
\author[1,3]{Anderson Janotti}
\author[1,3]{Xi Wang}
\author[1,2,3,*]{Chitraleema Chakraborty}

\affil[1]{Department of Materials Science and Engineering, University of Delaware, Newark, DE 19716, USA}
\affil[2]{Department of Physics and Astronomy, University of Delaware, Newark, DE 19716, USA}
\affil[3]{Quantum Science and Engineering Program, University of Delaware, Newark, DE 19716, USA}
\affil[4]{National Institute for Materials Science (NIMS), Tsukuba, Ibaraki 305-0044, Japan}
\affil[5]{Center for Computational Quantum Physics, Flatiron Institute, 162 5th Avenue, New York, New York 10010, USA} 
\affil[6]{Department of Physics and Astronomy, Stony Brook University, Stony Brook, New York, 11794-3800, USA}
\affil[*]{Corresponding author: cchakrab@udel.edu}

\date{} % Leave empty to remove date or use \today

\begin{document}

\maketitle
\vspace{-4em} % Adjust this number to pull the abstract up as much as you need

%TC: ignore
\begin{abstract}
\bfseries
Optical polarization is a key degree of freedom in quantum photonic and sensing technologies, enabling efficient light--matter coupling and directional emission. However, in defect ensemble-based quantum systems, orientational averaging across many defects suppresses optical anisotropy and eliminates a deterministic polarization axis. Here, symmetry engineering is introduced as a strategy to restore collective optical anisotropy in negatively charged boron vacancy (\VBm) ensembles hosted in hexagonal boron nitride (h-BN). By imposing anisotropic in-plane tensile strain through lithographically defined nano-ridge arrays, in-plane symmetry is broken and the ensemble emission dipole is aligned with the ridge-defined strain symmetry axis. Polarization-resolved photoluminescence reveals that the polarization visibility scales with the magnitude of strain anisotropy, while the emission orientation rigidly follows the engineered symmetry axis across devices with varying strain direction. This deterministic alignment is independent of crystal orientation, establishing the optical foundation for polarization-defined ensemble spin readout and programmable photonic integration in two-dimensional quantum materials.
\end{abstract}
%TC: endignore

\vspace{0.5cm} 

\section*{1 \textbar\quad Introduction}
Optically active quantum defects in solid-state materials offer compelling opportunities for integrated quantum photonics and sensing, enabling the control of spin and optical degrees of freedom within scalable device platforms \cite{wolfowicz2021quantum,aharonovich2016solid,kim2025quantum}. They possess well-defined optical properties and are central to discrete variable quantum technologies \cite{andersen2015hybrid}. In quantum sensing, ensemble quantum emitters leverage large numbers of emitters to enhance total photoluminescence (PL) and statistical robustness over single defect sensing. In the shot-noise-limited regime, sensitivity scales as $1/\sqrt{N}$ with emitter number $N$, as demonstrated in diamond magnetometry \cite{Taylor2008HighSensitivity}. However, at the ensemble level, orientational averaging erases optical anisotropy, eliminating a deterministic polarization axis and limiting efficient coupling to polarization-sensitive photonic modes, where coupling depends on the alignment between emitter dipole orientation and the guided mode polarization \cite{peyskens2019integration}, while also degrading spin-based sensing performance \cite{acosta2009diamonds,rondin2014magnetometry,lesik2014perfect}. For example, NV$^{-}$ centers in diamond, which remain the benchmark solid-state defect platform \cite{rovny2024nanoscale,barry2020sensitivity,du2024single}, occupy four distinct orientations within the lattice, such that an applied magnetic field produces non-equivalent magnetic moments across the ensemble. As a result, a given microwave frequency is resonant with only a subset of NV$^{-}$ centers, while the remaining orientations contribute background photoluminescence without undergoing spin transitions, reducing spin contrast and magnetic sensitivity \cite{acosta2009diamonds,rondin2014magnetometry,lesik2014perfect}.

Significant efforts have been devoted to controlling defect orientation in NV$^{-}$ centers. Preferential alignment can be achieved during growth through crystallographic engineering \cite{lesik2014perfect,michl2014perfect}, while post-growth approaches based on strain engineering \cite{karin2014alignment} can bias the thermodynamic stability of specific NV$^{-}$ orientations; however, reorientation typically requires high-temperature annealing to overcome the large activation energy. Ultimately, these strategies remain constrained by the rigid three-dimensional lattice, limiting post-fabrication tunability of defect orientation.

Two-dimensional materials provide a fundamentally different opportunity to overcome these limitations. Their atomically thin geometry enables planar integration, mechanical deformability, and externally applied symmetry-breaking perturbations that are inaccessible in bulk hosts \cite{azzam2021prospects,reserbat2021quantum,kianinia2022quantum,murray2026quantifying,shaikh2025magnetic,chakraborty2024flatland}. Hexagonal boron nitride (h-BN) has emerged as a leading platform for solid-state quantum photonics \cite{aharonovich2016solid,tran2016quantum}, owing to its wide bandgap \cite{cassabois2016hexagonal}, chemical robustness, and wafer-scale compatibility \cite{wang2019epitaxial,chang2017growth}. Within h-BN, the negatively charged boron vacancy (\VBm) defect exhibits room-temperature optical spin properties and stability \cite{gottscholl2020initialization,gottscholl2021room,liang2023high,sarkar2023identifying,kianinia2020generation}, and has been explored for quantum sensing \cite{healey2023quantum,lyu2022strain,mu2026spin,zabelotsky2023creation}. Despite these advantages, boron vacancy defect ensembles in h-BN face the same fundamental limitation as NV$^{-}$ centers in diamond: because optically detected magnetic resonance (ODMR) transitions in \VBm\ are primarily sensitive to magnetic-field projections along the spin quantization axis \cite{zheng2024magnetic}, orientational disorder reduces ensemble optical polarization contrast or visibility and also limits magnetic sensing performance \cite{wen2025alignmentfree}. Overcoming this limitation requires establishing a common optical symmetry axis across the ensemble, as the optical response of the \VBm defect in h-BN is governed by symmetry. In its pristine $D_{3h}$ configuration, the defect is subject to symmetry-locked optical transitions with restrictive selection rules that limit the available radiative pathways \cite{ivady2020ab}. 

This symmetry dependence provides a direct opportunity for external control: anisotropic strain can lower the local symmetry of the defect and thereby modify its optical selection rules. More broadly, strain has been used to control the polarization of localized emitters in two-dimensional materials, particularly in monolayer transition-metal dichalcogenides. Wang et al. demonstrated highly linearly polarized quasi-one-dimensional localized excitons along substrate protrusion edges in monolayer WSe$_2$, with the polarization arising from strain-induced confinement and valley hybridization \cite{wang2021highly}, while So et al. achieved position- and polarization-controlled single-photon emission in monolayer WSe$_2$ using nanogap-defined tensile strain \cite{so2021polarization}. More recently, Paralikis et al. demonstrated deterministic polarization control of individual WSe$_2$ emitters using tailored nanopillar geometries, aligning the emission dipole with lithographically defined nanowrinkles \cite{paralikis2024tailoring} and Moon et al. reported linearly polarized room-temperature emission arising from multiple strain-confined localized excitonic states in monolayer MoS$_2$\cite{moon2026highly}. These prior demonstrations involve localized excitonic states, with their polarization arising from strain-induced localization, confinement, and associated excitonic or valley physics, and are therefore distinct from the defect-symmetry mechanism relevant to \VBm, where anisotropic strain is expected to directly modify the local symmetry and optical selection rules of an atomic point defect.

Studies of individual quantum emitters in h-BN have demonstrated strong polarization visibility \cite{horder2024near} and discrete polarization axes \cite{kumar2024polarization}, and have further shown that local strain can perturb their optical transitions \cite{yim2020polarization,grosso2017tunable}. However, these results involve diverse or unidentified defect species and remain inherently stochastic. Because such effects are confined to isolated emitters, they do not provide a route toward ensemble-wide control. To date, research on \VBm\ centers has primarily focused on defect formation and energetics \cite{kianinia2020generation,liang2023high}, spin and charge-state engineering \cite{gale2023manipulating,fraunie2025charge}, nuclear-spin initialization and coherent control \cite{gao2022nuclear}, and strain-induced modifications to optical and spin response \cite{mu2026spin,geng2026deformation}, leaving deterministic control of ensemble optical polarization as an outstanding challenge. 

Here we demonstrate that engineered anisotropic in-plane tensile strain enables deterministic symmetry control in \VBm\ ensembles in h-BN. By suspending few-layer h-BN over silicon nano-ridge arrays, we generate spatially controlled anisotropic strain that defines a lithographically programmable symmetry axis. Complementary first-principles calculations show that anisotropic strain lifts the degeneracy of the bright in-plane defect transition, thereby enhancing the optical brightness along with establishing a principal optical axis. This strain-defined axis directly sets the ensemble emission orientation, enabling polarization control by design. We find that polarization visibility scales with strain magnitude, while polarization direction is set by strain symmetry, establishing amplitude and direction as independently tunable degrees of freedom. This approach provides a scalable route to deterministic ensemble polarization control, enabling polarization-matched photonic integration and establishing a platform for symmetry-defined ensemble quantum sensing.

\section*{2 \textbar\quad Results and discussion}

%TC:ignore

\begin{figure}[!htb]
\centering
\includegraphics[width=0.9\textwidth, trim=0cm 3cm 0cm 2cm, clip]{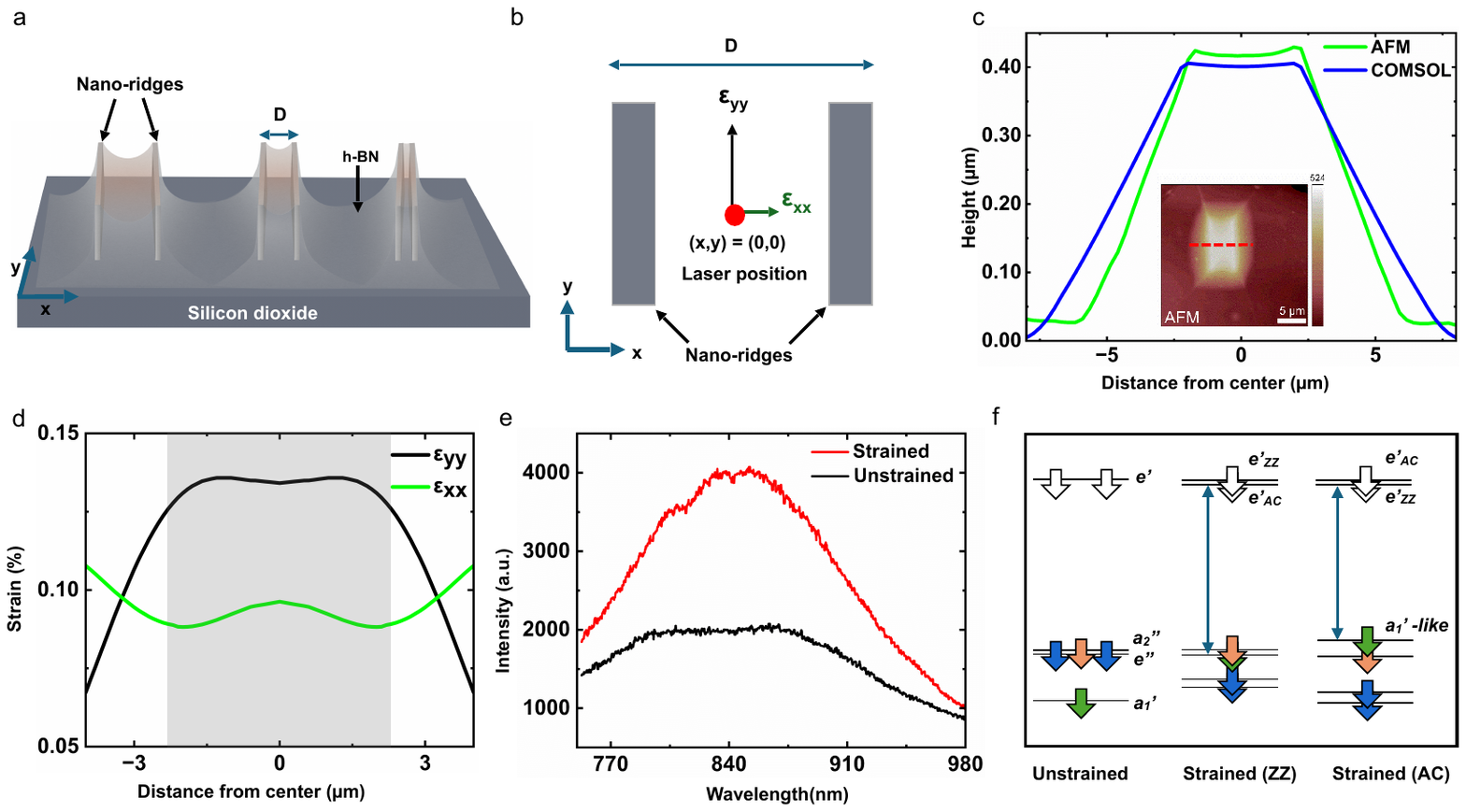}

\caption{\textbf{Device geometry, anisotropic strain generation, and strain-enhanced photoluminescence in suspended h-BN.} 
\textbf{a}, Schematic of multilayer h-BN suspended across silicon nano-ridges, generating anisotropic in-plane tensile strain. 
\textbf{b}, Coordinate definition and ridge spacing $D$ (2.5--8.1~\textmu m). The $y$ axis is parallel to the ridges and the $x$ axis corresponds to the ridge-to-ridge direction. The red dot marks the optical measurement position at the center of the suspended region. 
\textbf{c}, Height profiles across a representative suspended bridge from AFM measurements and finite-element simulations. Inset: AFM topography with the linecut used for profile extraction. 
\textbf{d}, Simulated in-plane strain components $\varepsilon_{xx}$ and $\varepsilon_{yy}$ along the $x$ direction between adjacent ridges for a representative device, demonstrating an anisotropic strain distribution within the suspended region (shaded). 
\textbf{e}, Representative photoluminescence spectra from strained (red) and unstrained (black) regions acquired under identical excitation conditions, showing enhanced emission intensity under applied strain. 
\textbf{f}, Calculated spin-down single-particle defect levels of $V_B^-$ for unstrained and anisotropically strained h-BN. In the unstrained defect, the $e' \rightarrow a_1'$ in-plane transition involves a degenerate $e'$ excited-state doublet, resulting in no preferred in-plane optical axis. Anisotropic strain along the zigzag (ZZ) or armchair (AC) direction lifts the $e'$ degeneracy, producing nondegenerate branches with distinct in-plane dipole character and modified optical selection rules (blue arrows). Under AC strain, the ground-state manifold additionally acquires $a_1'$-like character, reflecting symmetry lowering at both the excited- and ground-state levels.}
\label{fig:strain_device}
\end{figure}
%TC:endignore 
Suspended h-BN membranes were formed by transferring multilayer flakes with thicknesses of 37–60 nm onto arrays of silicon nano-ridges fabricated on a silicon-on-insulator substrate (Fig.~1a). This geometry induces anisotropic in-plane tensile strain within the suspended regions, with both in-plane components $\varepsilon_{xx}$ and $\varepsilon_{yy}$ remaining finite due to elastic coupling. The magnitude and direction of the strain field are defined by the ridge spacing $D$ (2.5–8.1~$\mu$m) and ridge orientation, respectively. During stamping, the membrane conforms to the ridge topography and is stretched in both in-plane directions. After removal of the applied force, adhesion to the substrate fixes the bridge endpoints, defining the final equilibrium strain state.

Throughout this work, the direction parallel to the ridges defines the $y$ axis, while the ridge-to-ridge direction corresponds to the orthogonal $x$ axis (Fig.~1b). The residual strain field after transfer is governed by adhesion and post-transfer membrane relaxation, allowing tensile strain to develop along both in-plane directions.  The suspended bridge is anchored at the ridge contact points along $x$, constraining tensile strain in that direction after removal of the stamping force. In contrast, no equivalent ridge constraint exists along the ridge-parallel ($y$) direction, allowing the $y$-directed strain response to vary more strongly with ridge spacing. Consequently, changing the ridge separation preferentially modulates $\varepsilon_{yy}$ relative to the more strongly constrained $\varepsilon_{xx}$, providing differential control of the in-plane strain anisotropy.

 Atomic force microscopy (AFM) topography measurements combined with finite-element simulations were used to quantify the deformation and resulting strain distribution (see Supplementary Note~1 for details). Figure~1c shows close agreement between AFM- and COMSOL-extracted height profiles across a suspended region, consistent with a quantitative predictive framework for nanostructure-induced strain in suspended van der Waals membranes \cite{murray2026quantifying}. The simulated in-plane strain components for a representative device ($D = 4.1~\mu$m; Fig.~1d) reveal a strongly anisotropic strain field within the suspended membrane. 
The $\varepsilon_{yy}$-dominant regime arises from the finite two-dimensional ridge geometry and surrounding membrane adhesion, with geometry-controlled simulations confirming that the relative magnitudes of $\varepsilon_{xx}$ and $\varepsilon_{yy}$ depend on ridge aspect ratio and boundary conditions (Supplementary Note~1).
 Varying the ridge spacing predominantly modulates the ridge-parallel component $\varepsilon_{yy}$, whereas $\varepsilon_{xx}$ exhibits only modest variation across devices and shows no systematic correlation with polarization visibility. Ridge spacing, therefore, enables tuning of strain anisotropy primarily through $\varepsilon_{yy}$, which serves as the principal engineered strain parameter in the analysis that follows. The complete evolution of both $\varepsilon_{xx}$ and $\varepsilon_{yy}$ with ridge spacing is provided in the Supplementary Note 3.

To probe the optical response under anisotropic strain, \VBm defect ensembles were generated uniformly across the h-BN flakes by Xe$^{+}$ ion irradiation, yielding high-density emitters in both strained (suspended) and unstrained (substrate-supported) regions. Ensemble photoluminescence (PL) spans the 750–1000~nm range centered around $\sim$850~nm, consistent with prior reports of \VBm emission \cite{gottscholl2020initialization,gottscholl2021room}. 
The presence of the spin-active V$_\mathrm{B}^{-}$ centers was further confirmed by room-temperature optically detected magnetic resonance (ODMR) measurements on a representative unstrained, Xe$^{+}$-irradiated h-BN sample with emission in the same spectral band (Supplementary Figure~S5 and Supplementary Note~2).
 Figure 1e compares representative spectra acquired under identical excitation and collection conditions from strained and unstrained regions, revealing a pronounced enhancement of PL emission intensity under applied strain.
 Strain-associated enhancement of defect emission in h-BN has also been reported in previous studies, both under controlled strain and in deformation regimes involving additional local perturbations.\cite{yang2022spin,geng2026deformation}

We interpret the observed PL enhancement as a consequence of strain-induced symmetry lowering and the associated relaxation of optical selection rules. Throughout this work, lower-case irreducible-representation labels ($a_{1}'$, $e'$, $\dots$) denote single-particle states, while upper-case labels ($^{3}A_{2}'$, $^{3}E''$, $\dots$) denote the corresponding many-body states. In the unstrained $D_{3h}$ defect, the lowest-energy emissive many-body transition $^{3}E'' \rightarrow {}^{3}A_{2}'$ is in-plane dipole-forbidden; the only in-plane dipole-allowed transition between the ground state and the lowest-lying triplet manifold is $^{3}A_{2}' \rightarrow {}^{3}E'$, which is populated only transiently before nonradiative relaxation to the lower-lying $^{3}E''$ state.  Consequently, the ensemble-averaged radiative response is intrinsically weak and lacks a preferred in-plane axis. Anisotropic in-plane strain reduces the local symmetry to $C_{2v}$, splitting the $E''$ manifold into $A_{2} \oplus B_{2}$ and partially lifting the dipole forbiddance: one component of the previously forbidden transition becomes dipole-allowed under symmetry lowering (see Supplementary Note~5 for the full group-theoretical analysis). 
To examine this strain-induced response at the defect level, first-principles calculations were performed using a periodic bulk h-BN supercell under prescribed anisotropic strain.
 Our first-principles calculations corroborate this picture at the single-particle level (Fig.~1f), showing that anisotropic strain along either zigzag or armchair directions splits the doubly degenerate $e'$ manifold into two nondegenerate branches with distinct in-plane dipole characters. The single-particle picture therefore predicts both a strain-induced enhancement of \VBm\ PL as seen in Fig. 1e and the emergence of linearly polarized emission under anisotropic strain, confirmed in the following section using polarization-resolved spectroscopy.

%TC:ignore
\begin{figure}[!htb]
\centering
\includegraphics[width=1.0\textwidth, trim=0cm 3cm 0cm 2cm, clip]{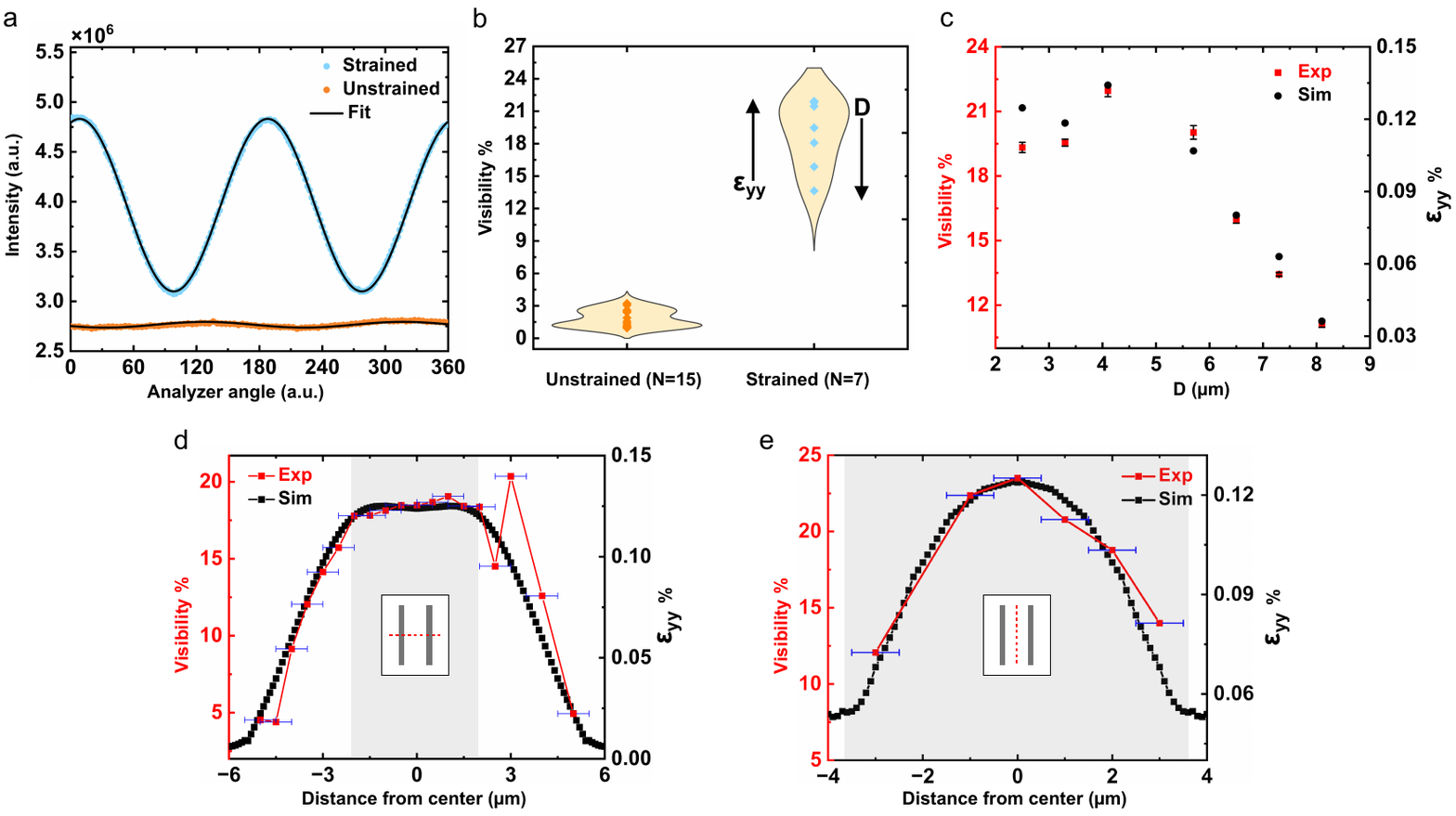}

\caption{\textbf{Optical mapping of anisotropic strain in \VBm ensembles.} 
\textbf{a}, Polarization-resolved PL intensity as a function of analyzer angle for strained and unstrained regions. Solid lines are fits to the dipole emission model. 
\textbf{b}, Statistical distribution of polarization visibility for unstrained ($N = 15$) and strained ($N = 7$) regions. The black arrows indicate the changes in magnitude of the strain component $\varepsilon_{yy}$ and the ridge spacing $D$. 
\textbf{c}, Polarization visibility as a function of ridge spacing $D$, compared with simulated $\varepsilon_{yy}$. Error bars represent the standard deviation from four repeated measurements. 
\textbf{d,e}, Spatial variation of visibility measured along horizontal (\textbf{d}) and vertical (\textbf{e}) linecuts across a device with $D = 4.1~\mu$m. The grey shaded region denotes the suspended area used for analysis. Error bars reflect the $\pm 0.5~\mu$m positional resolution of the differential translation stage. Insets: device schematics indicating the corresponding linecut directions.}
\label{fig:strain_mapping}
\end{figure}
%TC:endignore 
To probe the emergence of strain-induced linearly polarized emission, polarization-resolved PL measurements were performed in unstrained and suspended regions with systematically engineered anisotropic strain defined by the ridge spacing $D$.

Measurements were acquired at the centers of suspended membranes, where anisotropic tensile strain is maximal. The PL intensity, integrated over the ensemble emission band, was recorded as a function of analyzer angle under fixed linearly polarized excitation. Strained regions exhibit pronounced $180^\circ$-periodic angular modulation, indicative of linearly polarized ensemble emission, whereas unstrained regions show very limited angular dependence, consistent with orientational averaging over randomly aligned defect dipoles (Fig.~2a). The polarization response is well described by
\[
I(\alpha) = A \cos\!\bigl[2(\alpha - \theta)\bigr] + B,
\]
where $\alpha$ is the analyzer angle, $\theta$ denotes the ensemble-averaged emission polarization orientation, and $A$ and $B$ are fitting parameters corresponding to the modulation amplitude and polarization-averaged intensity, respectively.

In this framework, the polarization visibility serves as an optical metric of anisotropic in-plane strain in h-BN. The polarization visibility,
\[
V = \frac{I_{\max} - I_{\min}}{I_{\max} + I_{\min}},
\]
was extracted from 15 unstrained locations and 7 suspended regions spanning ridge spacings of $D = 2.5$--$8.1~\mu$m. As shown in Fig.~2b, the visibility increases systematically with decreasing $D$, reaching a maximum of 22\% before saturating. This saturation coincides with a reduction in the simulated $\varepsilon_{yy}$ component at small $D$ (Fig.~2c), suggesting that the observed visibility ceiling primarily reflects the strain distribution imposed by the current device geometry rather than an intrinsic limit of the polarization response. Incomplete strain transfer in multilayer suspended membranes, where interlayer shear compliance enables partial relaxation, may additionally contribute to this behavior \cite{kumar2016limits}. Such effects are not captured by continuum finite element simulations.

Figure~2c directly compares the experimentally extracted polarization visibility with finite-element simulations of the ridge-parallel in-plane strain component $\varepsilon_{yy}$ as a function of ridge spacing. The strong correspondence identifies $\varepsilon_{yy}$ as the principal strain component governing the magnitude of ensemble optical anisotropy within the experimentally accessible range. Within this regime, the ensemble polarization response follows $\varepsilon_{yy}$ and shows no systematic correlation with $\varepsilon_{xx}$ (see Supplementary Note 3).

The spatial sensitivity of this polarization-based readout is further demonstrated through linecut measurements for the ridge pair with $D = 4.1\,\mu$m, corresponding to the maximum visibility in Fig.~2c. As shown in Fig.~2d and e, the ensemble-averaged polarization visibility closely tracks the simulated spatial variation of the longitudinal strain component, $\varepsilon_{yy}$, within the suspended region. 
We note that, as the polarization visibility is normalized, effects arising from the suspended structure that may affect both polarization components equally, such as position-dependent Fabry--Pérot intensity modulation, cancel in the normalization and therefore do not contribute to the measured visibility.
 Linecuts taken along both horizontal and vertical directions capture the peak of the strain profile near the bridge center and its gradual relaxation toward the ridge edges. The agreement between experiment and simulation along these orthogonal directions further demonstrates that the polarization signal accurately tracks the local strain distribution across the suspended bridge. Together, these results establish ensemble emission polarization as a spatially resolved, non-invasive optical probe of anisotropic tensile strain in h-BN.

%TC:ignore
\begin{figure}[!htb]
\centering
\includegraphics[width=0.9\textwidth, trim=1cm 5cm 1cm 7cm, clip]{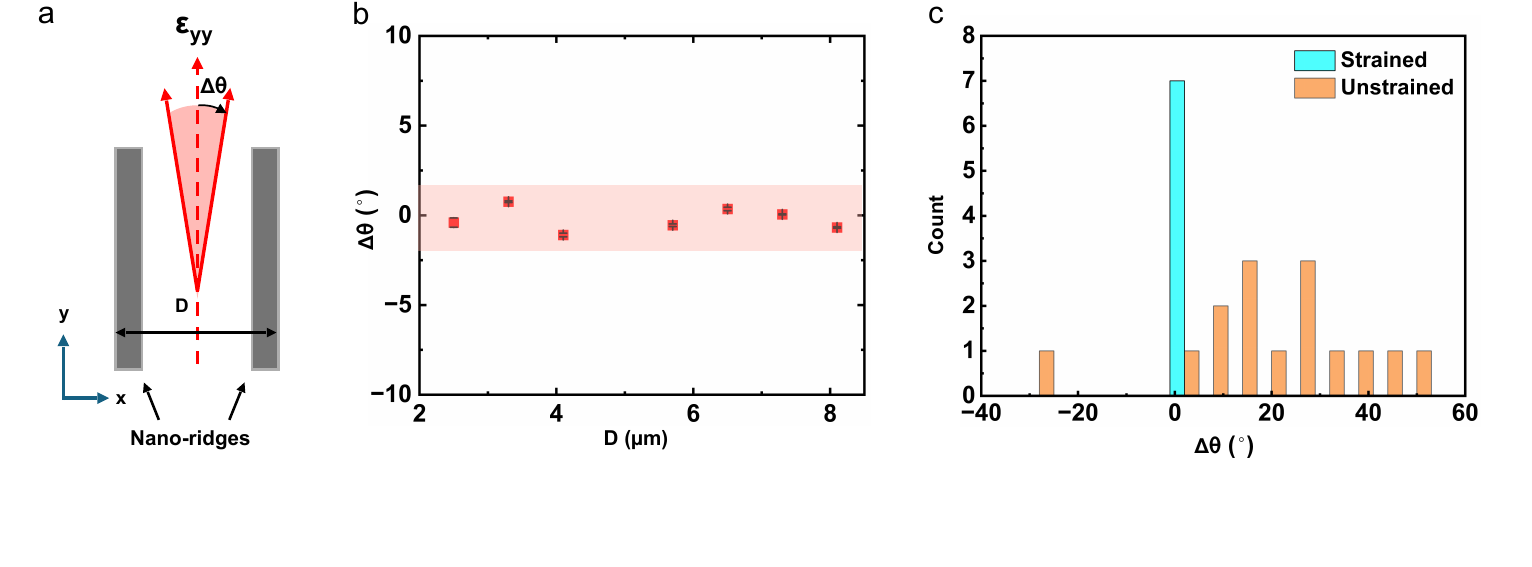}

\caption{\textbf{Strain-defined emission dipole orientation in \VBm ensembles.} 
\textbf{a}, Measurement geometry defining the ridge-parallel principal strain axis $\varepsilon_{yy}$ used as the reference for extracting the angular deviation $\Delta\theta$ between the ensemble polarization orientation and the strain axis. 
\textbf{b}, Angular deviation $\Delta\theta$ as a function of ridge spacing $D$. The deviation remains centered at zero within experimental uncertainty across all devices, indicating deterministic alignment of the emission dipole with the strain-defined symmetry axis. Error bars represent the standard deviation from four independent measurements at each spacing. 
\textbf{c}, Statistical distribution of $\Delta\theta$ for unstrained (orange) and strained (light blue) regions. Unstrained regions exhibit a broad angular distribution consistent with orientational averaging, whereas strained regions exhibit a narrow distribution centered at $\Delta\theta \approx 0^\circ$.}
\label{fig:dipole_orientation}
\end{figure}
%TC:endignore 
We next examine whether anisotropic strain can also deterministically set the orientation of the ensemble emission polarization. In unstrained regions, the degenerate $e'$ manifold supports orthogonal in-plane dipoles without a unique principal axis, so ensemble averaging suppresses a preferred polarization direction. By contrast, anisotropic tensile strain lifts the degeneracy of the in-plane transitions through symmetry lowering, thereby establishing a reproducible polarization axis defined by the strain symmetry.

The ensemble polarization angle $\theta$ is extracted from polarization-resolved measurements using a rotating half-wave plate and fixed analyzer in the collection arm. To quantify alignment with the engineered strain axis, we define the angular deviation $\Delta\theta = \theta - \phi$, where $\phi$ is the angle between the ridge orientation and a fixed laboratory axis.

Figure~3a illustrates the measurement geometry and the ridge-parallel strain axis $\varepsilon_{yy}$ that defines the reference frame for extracting $\Delta\theta$. As shown in Fig.~3b, $\Delta\theta$ remains centered at zero within experimental uncertainty across the full range of ridge spacings $D$ investigated, with no systematic dependence on $D$. Error bars represent the standard deviation from four independent measurements at each spacing. These results demonstrate that strain symmetry, not strain magnitude, governs the polarization axis.

The robustness of this strain-defined dipole locking is further illustrated in Fig.~3c, which compares the distributions of $\Delta\theta$ for strained (N = 7) and unstrained (N = 15) regions. While unstrained regions exhibit a broad angular distribution consistent with orientational averaging, strained regions exhibit a narrow distribution centered at $\Delta\theta \approx 0^\circ$. These results establish that anisotropic tensile strain not only enhances polarization visibility but also enforces a deterministic dipole orientation at the ensemble level.

Building on the strain-locked dipole orientation established above, we next demonstrate that the ensemble emission direction can be deterministically programmed through device geometry. As illustrated in the device schematic in Fig.~4a, the nano-ridge pairs are fabricated with a fixed spacing while their orientation $\phi$ is incrementally rotated relative to a fixed laboratory reference frame. Polarization-resolved measurements were first performed on a device with ridge spacing $D = 5.7~\mu\mathrm{m}$ (flake~1).

As $\phi$ is incrementally rotated, the polarization-resolved emission profiles shift with $\phi$ (Fig.~4b), demonstrating that the ensemble emission orientation is determined by the strain geometry imposed by the nano-ridges.
This rigid rotation is consistent with strain-induced symmetry lowering of the calculated $e' \rightarrow a_1'$ transitions shown in Fig.~1f, which lifts the degeneracy of in-plane optical transitions. Rotating the ridge orientation correspondingly rotates the preferred emission polarization axis in the laboratory frame, yielding $\theta \approx \phi$.

To verify that the observed behavior is not flake-specific, the experiment was repeated on a separately transferred h-BN flake on the same substrate with a different ridge spacing ($D = 3.1~\mu\mathrm{m}$, flake~2). The selected ridge spacings were chosen to produce comparable strain and polarization visibility, based on the trend in Fig.~2c. The extracted emission dipole orientations for both flakes are summarized in Fig.~4c, where the dipole angle $\theta$ closely follows the ridge orientation $\phi$ within experimental uncertainty. The solid line shows $\theta = \phi$ for visual reference, indicating rigid tracking of the engineered strain axis with no significant angular offset. Error bars represent the standard deviation obtained from four measurements. The close agreement between the two devices demonstrates that the dipole alignment is dictated by the engineered strain axis and is independent of the underlying crystal orientation of the host lattice.

%TC: ignore
\begin{figure}[!htb]
\centering
\includegraphics[width=0.9\textwidth, trim=0cm 3cm 0cm 0cm, clip]{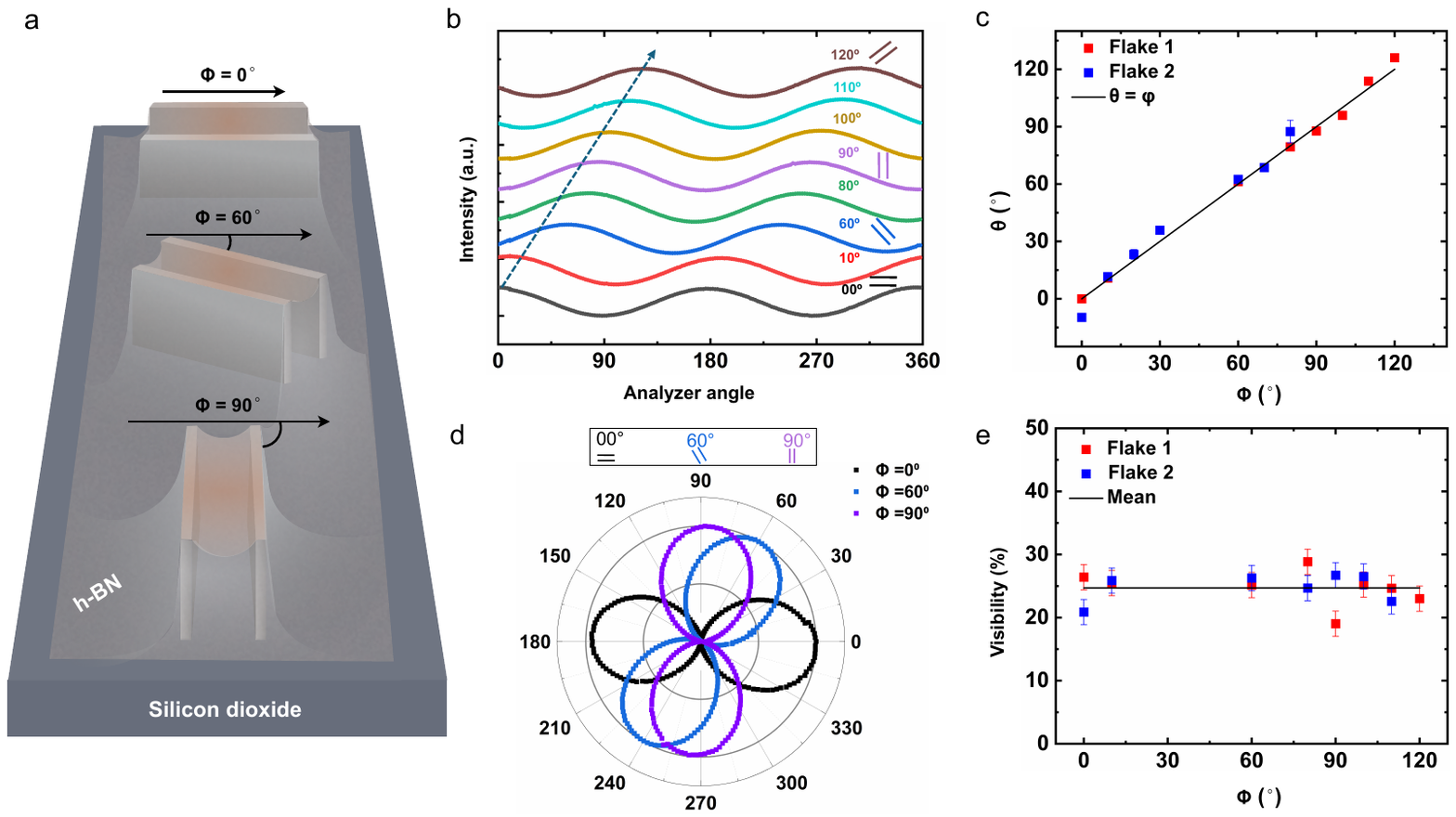}

\caption{\textbf{Programmable control of ensemble emission polarization via strain-axis engineering.}
\textbf{a,} Device schematic showing nano-ridge pairs with fixed spacing and incrementally rotated orientation $\phi$, defining the strain axis in the suspended h-BN membrane.
\textbf{b,} Polarization-resolved photoluminescence as a function of analyzer angle for devices with different ridge orientations $\phi$. Traces are vertically offset for clarity.
\textbf{c,} Extracted emission dipole angle $\theta$ versus ridge orientation $\phi$ for two independently transferred flakes (flake~1: $D = 5.7~\mu\mathrm{m}$; flake~2: $D = 3.1~\mu\mathrm{m}$). The solid line indicates $\theta = \phi$ for visual reference. Points represent averages of four measurements; error bars denote the standard deviation.
\textbf{d,} Polar plots of normalized polarization-resolved emission for representative ridge orientations ($\phi = 0^\circ$, $60^\circ$, and $90^\circ$), showing rotation of the emission dipole with the engineered strain axis. Data are obtained from flake~1.
\textbf{e,} Polarization visibility versus ridge orientation $\phi$ for both flakes. Mean visibility values are $24.7 \pm 2.8\%$ (flake~1) and $24.8 \pm 2.3\%$ (flake~2). The horizontal line at $24.7\%$ indicates the mean visibility.
}
\label{fig:polarization_control}

\end{figure}
%TC:endignore 

Polar plots for selected ridge orientations in flake 1 (Fig.~4d) provide a complementary visualization of this effect, showing rigid rotation of the emission lobe. Polarization-resolved emission maps across the full spectral range show the same rotation of the polarization axis with orthogonal ridge orientations (see Supplementary Note 4 and Fig.~S7).

Importantly, the polarization visibility remains largely unchanged across the full range of ridge rotation angles for both devices, as shown in Fig.~4e. The measured visibilities cluster around nearly identical mean values of $24.7 \pm 2.8\%$ for flake~1 and $24.8 \pm 2.3\%$ for flake~2, demonstrating that the polarization visibility remains consistent across independently transferred flakes as long as the strain magnitude remains comparable. The error bars represent the standard deviation of four independent measurements across each flake. A horizontal line at $24.7\%$ is included in Fig.~4e as a visual guide to highlight the near-constant visibility across ridge orientations. 
These results further confirm that rotating the nano-ridge orientation programs the emission polarization direction while preserving the polarization visibility within experimental uncertainty and material variability. More broadly, the experiments establish that the ridge geometry defines the orientation of the strain-selected in-plane optical axis, whereas the magnitude of the strain anisotropy determines the degree of ensemble polarization.

The comparable visibility values observed for flakes~1 and~2 are consistent with the strain-dependent measurements in Fig.~2c, where ridge spacings closest to those used here ($3.3$ and $5.7~\mu\mathrm{m}$) exhibit similar polarization visibility, and may additionally reflect device-to-device mechanical variations such as flake thickness, membrane tension, or adhesion.

Finally, we verify that the emission polarization $\theta$ is independent of excitation polarization. Measurements performed using orthogonal excitation polarizations yield identical emission angles (see Supplementary Note~4 and Supplementary Fig.~S8), demonstrating that the dipole orientation is passively set by device geometry rather than optical pumping conditions. Together, these results establish deterministic geometric control of ensemble emission polarization while preserving polarization visibility. This capability provides a pathway for routing photons in polarization-selective photonic architectures without external tuning mechanisms, relevant for integrated quantum photonic platforms where coupling efficiency depends on alignment between emitter dipole orientation and guided mode polarization.

\section*{3 \textbar\quad Conclusions}
In conclusion, we have demonstrated a strain-engineered platform that enables deterministic control and optical readout of ensemble emission polarization in \VBm centers hosted in hexagonal boron nitride.  
Anisotropic tensile strain introduced through device geometry simultaneously enhances polarization visibility and locks the ensemble emission dipole to the engineered strain axis by lifting the degeneracy of the bright in-plane defect optical transition and selecting a principal polarization direction.

A central outcome of this work is the identification of ensemble emission polarization as both a quantitative optical probe of anisotropic strain magnitude and a deterministic indicator of strain symmetry. Notably, this approach enables the detection and mapping of very small strain variations ($<0.2\%$), extending sensitivity to subtle anisotropic strain fields.
 The polarization visibility tracks the magnitude of the dominant strain component, while the emission dipole orientation directly encodes the orientation of the principal strain axis, enabling purely optical mapping of local strain fields. Together, these measurements establish independent control of polarization visibility and polarization direction through device design.

By imposing a common symmetry axis across defect ensembles, this approach enables controlled alignment of optical emission with polarization-sensitive photonic components and provides a pathway toward polarization-functional quantum emitters compatible with planar fabrication and large-area integration. These results establish strain-defined ensemble polarization as a scalable strategy for symmetry-engineered quantum materials, enabling programmable polarization control in integrated photonic and sensing architectures.

%TC: ignore
\section*{Data availability}
The data that support the findings of this study are available from the corresponding author upon reasonable request.
\section*{Acknowledgements}
The spectroscopy and AFM measurements were primarily supported by the National Science Foundation through OIA-2217786 and partially supported by NSF through the University of Delaware Materials Research Science and Engineering Center, DMR-2011824. DFT calculations were carried out using the computational resources from the National Energy Research Scientific Computing Center (NERSC), a Department of Energy Office of Science User Facility, through the NERSC award BES-ERCAP 0034471 (m5002), and the DARWIN computing system at the University of Delaware, which is supported by the NSF Grant No.~1919839.
This work was carried out in part at the Singh Center for Nanotechnology, supported by the NSF National Nanotechnology Coordinated Infrastructure Program under grant NNCI-2025608. AI, CM, and CC acknowledge support from the National Science Foundation (grant OIA-2217786). APV acknowledges partial support from the National Science Foundation (grant OIA-2217786). MHS, AH, and AJ acknowledge partial support from NSF MRSEC (DMR-2011824). R.M.S. acknowledges support from São Paulo Research Foundation (FAPESP) (grants 2025/06767-07, 2024/10789-5, 2021/06744-8). The authors thank Nazifa Tasnim Arony for preparing the schematic illustrations used in this work.

\section*{Author contributions}
AI, CC and EH conceived the idea, AI and CC designed the experiments. EH designed and fabricated the nanostructured substrates, performed the finite element analysis of the strain distributions, and carried out the 2D material transfer onto the nanoridge substrates. AI, MHS and CM performed the 2D material transfer on plain substrates. LM performed atomic force microscopy measurements. AI performed the FIB irradiation on the flakes. AI, SA, and AH developed the experimental setup. AI, AH, SA, and APV performed polarization-resolved PL spectroscopy. AI and LS fitted the data; AI, CC, EH, XW, PA, and AJ analyzed and interpreted the data. AI and CC prepared the initial manuscript. Theoretical calculations were performed by PA, RDS, CED, and AJ. MD, AJ, XW, and CC supervised the project. All authors participated in discussions and contributed to refining the manuscript. All authors have approved the final version of the manuscript.

\section*{Competing interests}
EH and XW are listed as inventors on a pending patent application filed by the University of Delaware related to this work. The remaining authors declare no competing interests.

\section*{Methods}
\textbf{Sample preparation}
Nanostructures were fabricated using commercially available silicon-on-insulator (SOI, WaferPro) wafers following a process adapted from previous work\cite{herrmann2023single}. Lift-off (LOR 3A) and electron beam resist (AR-P 6200.04) were sequentially spin-coated onto the SOI substrate (220\,nm Si device layer on 3\,$\mu$m thermal SiO$_2$) to form a bilayer resist stack of $\sim$250\,nm thickness. The resist was exposed using electron beam lithography (Raith EBPG 5200, 100\,kV) and developed in AR 600-546 developer followed by isopropyl alcohol.

A 25\,nm chromium (Cr) hard mask was deposited at 0.2\,\AA\,s$^{-1}$ using electron beam evaporation. After lift-off of excess Cr, nanostructures were defined by dry etching in an inductively coupled plasma (ICP) system (PlasmaTherm Versaline) using CHF$_3$ and C$_4$F$_8$ gases with flow rates of 30 and 40\,sccm, respectively. The RF power was set to 30\,W, ICP power to 300\,W, and chamber pressure to 8\,mTorr, with a total etch time of 5\,min. Following etching, the Cr hard mask was removed by immersing the samples in a commercially available Cr etchant at 40\,$^\circ$C for 1\,h.

Large-area hBN flakes with approximately uniform thickness were exfoliated from bulk crystals using PDMS (Gel-Pak\textsuperscript{\textregistered}) and immediately transferred onto the patterned substrates using a water-assisted transfer process\cite{li2015optoelectronic}. In this process, the flake remains on the PDMS from exfoliation to transfer, without selection based on thickness. A small volume of water was first picked up at the flake location on the PDMS and then deterministically stamped onto the target nanostructure. While the flake and substrate were in contact, the sample temperature was gradually increased to 95\,$^\circ$C to evaporate the water and release the flake from the PDMS. The sample was subsequently cooled to room temperature.

\textbf{Polarization-resolved photoluminescence (PL) analysis}
Polarization-resolved PL measurements were carried out at room temperature using a home-built confocal microscope excited with a continuous-wave 532~nm laser through a 40$\times$ objective (NA = 0.75). A linear polarizer was placed in the excitation path to ensure a well-defined linear polarization aligned vertically along the optical axis. To avoid polarization-dependent artifacts in the collection path, a glass window was used in place of a dichroic mirror for spectral routing combined with a 750 nm long pass filter. The polarization state of the emitted PL was analyzed using a rotatable half-wave plate in combination with a fixed linear polarizer. The half-wave plate was rotated in discrete angular steps using an automated mount (Thorlabs K10CR2), while the analyzer polarizer was fixed vertically along the optical axis (laboratory reference frame) to minimize polarization-dependent detection bias. Any setup-induced polarization offsets were calibrated through controlled benchtop measurements by placing a linear polarizer before the objective, applying known polarization rotations, and measuring the corresponding optical offset.

\textbf{Ion irradiation and defect generation}

Xe$^{+}$ ions were implanted into the h-BN flakes using a focused ion beam (Tescan S8252X Xe Plasma FIB). The implantation parameters were selected to generate a near-surface distribution of boron vacancies, with an effective depth of approximately 15--25~nm, consistent with what has been previously reported~\cite{kianinia2020generation}. Irradiation was performed at an acceleration energy of 30~keV and a beam current of 17~pA. The ion fluence was calculated from the irradiation time, beam current, and irradiated area following the approach outlined in ref~\cite{kianinia2020generation}.

For the strain-varying sample, irradiation was performed with a dwell time of approximately 1~$\mu$s per point, an exposure pitch of 25~nm, and 40 scans using a zig-zag scanning path over an approximately $284 \times 139~\mu$m$^{2}$ area, corresponding to an ion fluence of approximately $6.7 \times 10^{14}$~ions~cm$^{-2}$. For the direction-varying samples, irradiation was performed with a dwell time of approximately 1~$\mu$s per point, an exposure pitch of 35~nm, and 20 scans using a zig-zag scanning path over an approximately $289 \times 231~\mu$m$^{2}$ area, corresponding to an ion fluence of approximately $1.8 \times 10^{14}$~ions~cm$^{-2}$. No post-implantation annealing was performed.

\textbf{Atomic force microscopy}
Atomic force microscopy (AFM) measurements were performed using an Anasys NanoIR2 system to determine flake thickness and nanostructure geometry. Flake thickness was extracted from averaged height profiles and used as an input parameter for the COMSOL strain simulations. Full AFM scans of suspended regions were acquired to obtain complete height maps, enabling validation of simulated deformation profiles and optimization of applied boundary conditions. The heights of the bare nanostructures (ridges) were independently verified by AFM.

\bibliography{references}    % This must match your .bib filename
%TC:endignore 
\end{document}